# Crystal fields and Kondo effect: Magnetic susceptibility of Cerium ions in axial crystal fields

H.-U. Desgranges[1]

*Albert-Kusel-Str. 25, Celle, Germany*

The thermodynamic Bethe ansatz equations of the Coqblin-Schrieffer model have been solved numerically. The full N=6 (J=5/2) degeneracy of the Hund's rule ionic ground state of Ce is taken into account. Results for the temperature dependent magnetic susceptibility parallel and perpendicular to the crystal axis are presented. The deviations, due to the Kondo effect, to the non-interacting ion results are pointed out.

## 1. Introduction

The complex behavior of Cerium compounds is still attracting interest among experimentalist. The change of the effective spin-degeneracy N due to the interplay between Kondo and crystal field effects has been studied experimentally by investigating various Cerium based pseudo-ternary intermetallic substitution series [1].

The single ion Kondo model and its generalization to a N-fold degenerate ionic configuration, the SU(N) Coqblin-Schrieffer model [2], has been used successfully to describe the thermodynamic properties of dense Kondo systems [3-5]. However, for fitting the influence of crystal fields on the paramagnetic susceptibility experimentalists have had to resort to the text-book result for non-interacting ions [6-9].

A broad basis for comparison with experiments on the specific heat in zero magnetic field over the whole temperature range has recently been provided [10] and applied successfully [11]. The numerical solution of the thermodynamic Bethe ansatz equations for the N = 6 model (Cerium 3+ ions) with general crystal field configurations was achieved by a new high field / low temperature expansion to calculate the limiting values of the unknown functions. In a subsequent paper [12] the method was extended to the calculation of the magnetic susceptibility. However, this was restricted to the case of N = 4 that is applicable to Cerium in a temperature range in which the highest Kramers doublet may be neglected.

For the present work the method to calculate the magnetic susceptibility has been further developed to the N = 6 model with axial crystal fields that split the 6-fold degenerate ground state of the $Ce^{3+}$ ions into three doublets that are eigenstates of the total angular momentum operator $J_z$. Here results are presented for the case that the energy levels corresponding to the eigenstates $|\pm 1/2\rangle$, $|\pm 3/2\rangle$, $|\pm 5/2\rangle$ are sequenced in that order.

The new results allow for a quantitative comparison with experimental data with relevance to e.g. $CePt_5/Pt(111)$ [13].

## 2. Model and methods

The Bethe ansatz solution of the Coqblin-Schrieffer model [14, 15] was used by Schlottmann [16] to calculate the anisotropic magnetic susceptibility at zero temperature for the ionic crystal field Hamiltonian given by [16, 17]:

$$
\begin{aligned}
H_{ion} = b_2 O_2^0 &+ b_4 O_4^0 \\
&- g\,\mu_B S_z H_\parallel - g\,\mu_B S_x H_\perp
\end{aligned}
\tag{1}
$$

Here $O_2^0$ and $O_2^4$ denote the usual Stevens operators [18], and $H_\parallel$ and $H_\perp$ are the parallel and transversal projection of the magnetic field onto the crystal field axis, respectively.

For a parallel magnetic field the energy levels of the three doublets are given by:

$$
\begin{aligned}
E_{1/2} &= -8b_2 + 120b_4 \mp \tfrac{1}{2} g\mu_B H, \\
E_{3/2} &= -2b_2 - 180b_4 \mp \tfrac{3}{2} g\mu_B H, \\
E_{5/2} &= 10b_2 + 60b_4 \mp \tfrac{5}{2} g\mu_B H.
\end{aligned}
\tag{2}
$$

The calculation for a transversal magnetic field is done by second order perturbation theory [19] including van Vleck terms quadratic in H.

The calculation of thermodynamic properties follows the lines presented in the preceding publications [10, 12]. Details will be published elsewhere

E-mail address: H-Ulrich.Desgranges@nexgo.de
[1] Scientifically unaffiliated

[20]. The splittings between adjacent ionic energy levels $A_r \equiv E_{r+1} - E_r$, $1 \leq r \leq N$, serve as generalized fields determining the limiting values of the unknown functions of the infinite system of nonlinear integral equations. The energy levels according to eq. (2) have to be put into sequence such that $A_r \geq 0$. For reasons of brevity we restrict ourselves to the case $E_{1/2} < E_{3/2} < E_{5/2}$, region II in the notation of Schlottmann [16]. The extrapolation of our results to zero temperature has served as a check on our calculation for $A_2 = A_4$.

Temperature, magnetic, and crystal fields are scaled by the Kondo temperature in the absence of all fields $T_K(N=6)$. At low temperatures and large crystal fields the influence of the higher doublets may be neglected so that the thermodynamic properties are governed by an effective spin-1/2 system with an effective Kondo temperature $T_K(N=2)$ given by the relation

$$T_K(N=2) = \frac{(g\mu_B)^2}{2\pi \chi(T=0)}. \qquad (3)$$

For small temperatures compared with this low temperature scale the susceptibility decreases quadratically.

In this case eq. (3) may be used to determine $T_K(N=6)$ from the scaling relation $T_K^3(N=6) = C_4 A_2^2 T_K(N=2)$ where the numerical factor $C_4$ has been approximated by $C_4 = 17.08 + 17.05\, A_4/A_2$ [10].

For small values of the crystal fields the sixfold degeneracy at vanishing magnetic field is almost restored even at small temperatures. The qualitative behavior of the susceptibility is then the same as without crystal fields [21].

## 3. Results and conclusion

The numerical solution of the thermodynamic Bethe ansatz equations has been achieved for a number of crystal field splittings. The relative accuracy is expected to be better than 1%. All results are depicted in the form of the inverse magnetic susceptibility $\chi^{-1}$. The symbols indicate the finite temperature grid. All energies are measured in units of temperature ($k_B = 1$).

In Fig. 1. results for the Coqblin-Schrieffer model with axial crystal fields (solid lines) are compared with the corresponding curves for non-interacting ions [22] (dashed lines). A parallel magnetic field is considered exemplarily for energy splittings $A_2 / T_K(N=6) = 1$ while $A_4 / T_K(N=6)$ is varied.

The main point to show here is: While the principal outlook of the two sets of curves is similar there is a consistent quantitative discrepancy. This means that – if the Kondo effect plays a role in the low temperature physics of a certain compound – the fitting of the (inverse) susceptibility curves with the model of non-interacting ions cannot reliably be used to determine the crystal field splittings.

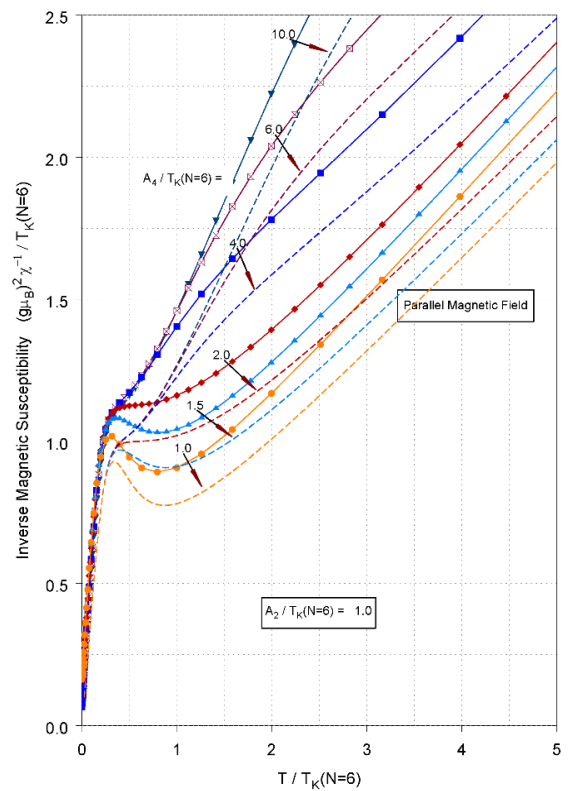

Fig. 1. (Color online) Inverse magnetic susceptibility $\chi^{-1}$ as function of temperature T scaled by the Kondo temperature in the absence of crystal fields compared with the corresponding non-interacting ion curves (dashed lines)

To facilitate a quantitative description of experimental results the numerical results for the Coqblin-Schrieffer model with axial crystal fields are depicted in Figure 2 for a fixed ratio of crystal field splittings $A_4 / A_2 = 1.0$, 2.0, and 4.0, respectively.

Here, the full symbols represent the results for a parallel magnetic field while the open symbols correspond to a transversal magnetic field. The dashed line shows the Coqblin-Schrieffer result without crystal fields [21].

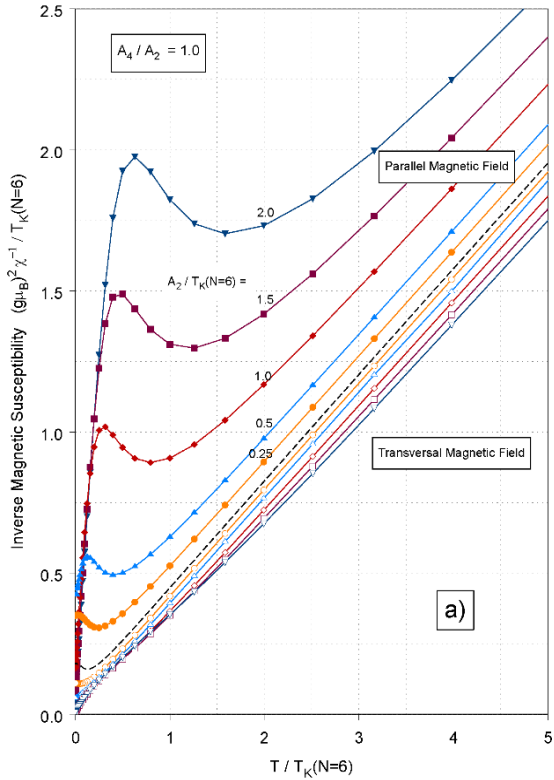

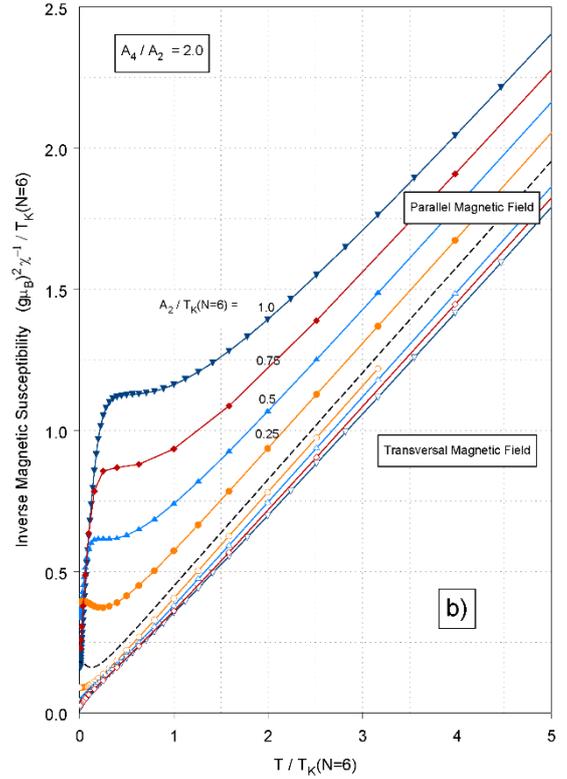

Fig. 2. (Color online) Inverse magnetic susceptibility $\chi^{-1}$ as function of temperature for parallel and transversal magnetic field. The ratio of crystal field splittings is kept fixed to a) 1.0, b) 2.0, and c) 4.0

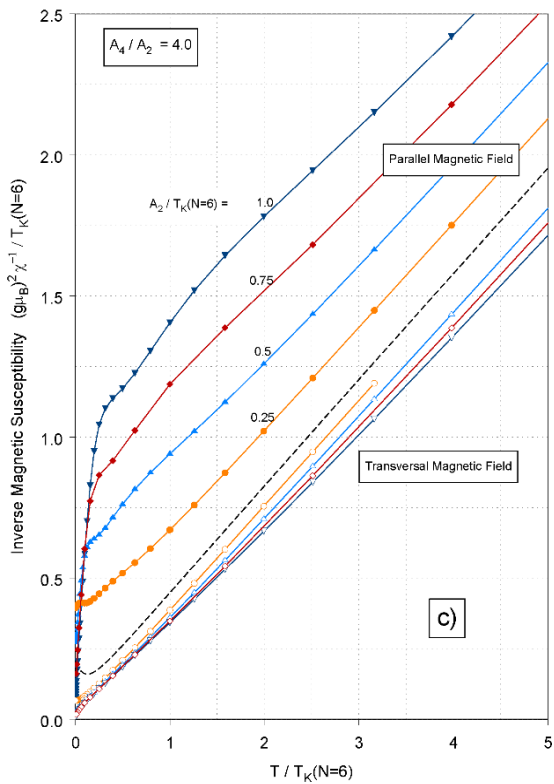

Unless the crystal field splitting is too small, the inverse susceptibility in a parallel magnetic field has its minimum at $T = 0$ and increases steeply at low temperatures. It then shows a pronounced peak for smaller values of $A_4 / A_2$. The peak decreases in height and turns into a plateau at intermediate values of $A_4 / A_2$. On increasing $A_4 / A_2$ further the plateau turns into a shoulder with an inflection point before the Curie-Weiss like high-temperature behavior (i.e. a straight line) is reached.

The inverse susceptibility in a transversal magnetic field in comparison does not display such distinct features. However, the curves bend to a straight line in about the same temperature range as the curves for the parallel magnetic field.

## 4. Summary

On the basis of a recently found new method the infinite set of coupled, nonlinear integral equations describing the thermodynamics of the N=6 Coqblin-Schrieffer model has been solved. Thereby the full degeneracy of the J=5/2 Hund's rule ionic ground state of Ce is taken into account. Results for the (inverse) anisotropic magnetic susceptibility for ions in an axial crystal field are presented to provide material for a quantitative analysis of experimental results [23].

The deviations, due to the Kondo effect, to the non-interacting ion results are pointed out.

## Acknowledgment

I thank Kai Fauth for providing some valuable references.